# Gender differences in research productivity: a bibliometric analysis of the Italian academic system[1] *

*Giovanni Abramo*[a,b**], *Ciriaco Andrea D'Angelo*[a] *and Alessandro Caprasecca*[a]

[a] Laboratory for Studies of Research and Technology Transfer
School of Engineering, Department of Management
University of Rome "Tor Vergata"

[b] Italian Research Council

**Abstract**

The literature dedicated to analysis of the difference in research productivity between the sexes tends to agree in indicating better performance for men. This study enters in the vein of work on the subject. Through bibliometric examination of the entire population of research personnel working in the scientific-technological disciplines of Italian university system, it confirms the presence of significant differences in productivity between men and women. But such differences result as being smaller than reported in a large part of the literature, confirming an ongoing tendency towards decline, and are also seen as more noticeable for quantitative performance indicators than other indicators. The gap between the sexes presents important sectorial specificities. In spite of the generally better performance of men, it can not be ignored that there are a significant number of scientific sectors in which the performance of women does not result as inferior.

**Keywords**

*Research productivity, gender differences, university, bibliometrics*

[1] Abramo, G., D'Angelo, C.A., Caprasecca, A. (2009). Gender differences in research productivity: a bibliometric analysis of the Italian academic system. *Scientometrics*, 79(3), 517-539. DOI: 10.1007/s11192-007-2046-8
* Authors are deeply indebted to Giorgia Barbetta for her invaluable support in data elaboration.
** Corresponding author, Università degli Studi di Roma "Tor Vergata", Facoltà di Ingegneria, Dipartimento di Ingegneria dell'Impresa, Via del Politecnico 1, 00133 Roma, ITALIA; Tel. +39 06 72597362, Fax +39 06 72597305, abramo@disp.uniroma2.it

# 1. Introduction

The study of differences in productivity between men and women employed in the scientific world has always attracted interest from a wide range of observers. It feeds a lively debate that follows at least two themes: psycho-cognitive and sociological. In particular, during the past two decades, the issues of gender differences in cognitive abilities has been addressed by numerous meta-analysis studies of verbal (Hyde and Linn, 1988), spatial (Linne and Peterson, 1985, Voyer et al., 1995) and mathematical abilities (Hyde et al., 1990). Such studies do not indicate a substantial differentiation in abilities between men and women but they do offer a characterization by typology of ability and context of application. In addition to this issue, for the sciences in particular, and especially in the world of research, the feminine presence still seems highly limited and relegated to marginal roles (UE, 2006):

- Women represent only one sixth of research workers in the private sector and one third of the entire community of academic staff, though their representation has increased over time.
- Regarding the composition of academic staff, women tend to be concentrated in inferior roles. There is only one woman for every 3.5 men in the top academic ranks.
- In the scientific committees appointed by the European Community the proportion of women is about 20%, but the leadership of these committees is entrusted to a woman in only 10% of cases.

Such statistics have stimulated studies of a sociological character to identify and suggest potential interventions to effectively balance the situation, commencing from the possible causes: the smaller number of women entering the field, unequal opportunity and sexual discrimination, or lesser performance with respect to men.

In this last area, one of the most consistent findings in the literature on research productivity is that women tend to have somewhat lower publication rates than men (Lee and Bozeman, 2005). The lesser productivity of females has been established in tens of studies of diverse countries and disciplines, spanning decades and using a wide variety of measures (Cole and Zuckerman, 1985, Fox, 193, Long 1987). Indications are that the difference in average productivity between the genders is also accompanied by a diverse distribution of the research product. Le Moine (1992) shows that the concentration of women among researchers who publish a single article is greater than for men, while their representation among "star" scientists is less. Cole and Zuckerman (1984) neatly label these gender differences as the "productivity puzzle", although science sociologists retain that the puzzle exists only for those who refuse to recognize the impact of sociological determinants.

Zainab's review of the studies on the subject identifies, among others, certain classes of personal variables that are significantly correlated to productivity of scientists. It results that the difference in scientific performance between men and women is significant, but also emerges that such differences lessen over time (Cole and Zuckerman, 1984; Xie and Shauman, 1998; Leahey, 2006), and can in part be traced to factors other than gender, such as level of specialization (Leahey, 2006) or academic position. Differences between the sexes in the early stages of career seem to be more visible (Xie and Shauman, 1998).

Within the vein of these investigations, the present study is intended to provide, for the first time, the evidence from the Italian academic research system. It proposes to

examine:
- Whether there are differences in research productivity between men and women;
- If such differences can be identified in all the evaluation parameters for scientific performance
- If such differences are general or present sectorial specificities;
- If such differences remain more or less constant or vary significantly with level of employment.

The current study is not intended to investigate, in this phase, the causes of the differences encountered, but the authors will indicate further investigations that the findings could suggest.

The work here is unique with respect to the international state of the art in at least two aspects. Firstly, for the field of observation − studies in the existing literature have been based only on samples of the population of interest and generally focalize either on very restricted disciplinary sectors or on single institutions. Instead, the study proposed here refers to the entire technological-scientific population of Italian universities, consisting of approximately 33,000 research scientists. Secondly, for the manner of comparing individual performance − each scientist has been classified by role and scientific field of specialization, with the aim of limiting the inevitable distortions in productivity due to non-homogeneity of gender distribution among roles and scientific sectors (see Abramo and D'Angelo, 2007). The research products are observed as the scientific publications in international journals recorded by the Thomson Scientific Science Citation Index (SCI™) during period 2001 to 2003. The analysis based on the whole population of academic research staff, avoids problems in robustness and significance of inferential analyses. It further presents the undeniable advantage of objectivity and homogeneity of source data, not always found in examinations based on questionnaires.

The work is articulated as follows. In section 2, the determinants of scientific production are analyzed. Section 3 presents the analytical model, the field of observation and the post-codification procedure used to construct the data set. Section 4, in addition to a brief presentation of some synthetic data relative to the field of observation, reports the investigation results with reference to the research questions that originally inspired the study. The paper closes with some further considerations by the authors.

## 2. Factors affecting scientific production

Research activities resemble a type of input-output process (Moravcsik, 1985), in which the inputs consist of human and financial resources, while outputs have a more complex character, of both tangible (publications, patents, conference presentations, etc.) and intangible nature (personal knowledge, consulting activity, etc.). The outputs most commonly used to evaluate results of research in science and technology are the scientists' publications in specialized journals, the form *par excellence* to communicate the results of their research to the community of reference. Through this medium, scholars obtain the recognition of their peers, a determining factor for further funding searches and career progression (Ramsden, 1994).

In comparing products of research work between any two individuals and particularly between the sexes, it is necessary to filter the effects of all factors other than individual merit that may affect individual performance in a direct or indirect manner. Zainab (1999) groups determinants of scientific performance in two categories: personal and environmental. The following determinants are noted among the first category:

- *Gender:* studies have revealed higher productivity among males, both in analyzing specific sectors and observing specific research institutions over an extended period of time (Fox, 2005; Stack, 2004; Xie and Shauman, 2004; Prpic, 2002).
- *Age:* certain studies seem to show the existence of a peak in productivity in the years approaching age 40 and the years soon afterwards, followed by a constant decline with advancing age (Fox, 1983). Investigation of the scientists at the National Research Council of Italy seems to confirm these findings (Bonaccorsi and Daraio, 2003).
- *Marriage:* Almost all studies agree on the positive effect of marriage on the scientific fertility of researchers, but certain studies (Prpic, 2002) show that men receive the greater share of the benefit due to the presence of a spouse. Fox (2005) shows that unmarried men are the least productive of all. Among women, those who are married, and particularly those married for the second or third time, have a higher level of productivity.
- *Children:* The results from investigations of the impact of children on productivity are not always simple to align. According to Fox (2005), the presence of children, especially of preschool age, increases productivity among both genders. Evidently children can motivate scientists to work harder, enabling them to provide a higher standard of living for their offspring. Women with preschool children are found to be especially efficient, particularly in their allocations of time. However, in a study of a much larger sample, Stack (2004) shows that women with preschool aged children publish less than other women. Obviously, the time, energy, and money devoted to child-rearing can reduce research productivity. In any case, men with children continue to be more productive than women with children (Prpic, 2002).
- *Level of specialization:* increases in professional specialization seem to have a positive of influence on a scientist's research productivity. Some studies illustrate that women tend to specialize less than men, which results to the detriment of their productivity (Leahey, 2006).

Certain structural and environmental factors can also be noted:

- *Academic rank (role):* many studies illustrate a correlation between academic rank and a scientist's productivity. In a study sample of American academics, Blackburn et al. (1978) show that full professors publish at a higher average rate than associate professors and research staff. Dickson (1983) and Kyvik (1990) have illustrated the same effect of professional role on scientific productivity in their respective studies of Canadian and Norwegian universities.
- *Teaching load:* in universities, research and teaching activities accompany each other. Certain studies of performance evaluation and gender show that men obtain better performance in research while women seem to excel in (and favour) teaching activities (Gander, 1999). In confirmation of this thesis, Xie and Shauman (2003) reveal that even though the difference in teaching load between the sexes is on the decline, women continue to favour teaching activity more than men and thus, on average, devote a lesser portion of time to research.
- *Prestige of the institution or department of affiliation:* certain studies illustrate that productivity may be a function of the researcher's institute of affiliation. The presence of "illustrious" colleagues has a positive effect on the productivity of the other researchers. Further, this effect is seen as more notable among lower level researchers (post-doctoral student, research associate, teaching fellow) and continually weaker as careers advance. But the direction of the cause-effect relationship is unclear, as to whether it is the better university staff teams that draw the most brilliant minds or vice versa (Fox, 1983).

In light of the state of the art, for this study the authors chose to take in consideration all the relevant factors cited above, either directly or indirectly, while also working with a field of observation much wider and more representative than those of the preceding studies. The following section of this paper proceeds with a description of the methodological choices, highlighting their relevance with respect to the present limits in the state of the art.

## 3. Analytical model

For the scope of this study, two variables among those indicated by current literature were taken into particular consideration: academic role and gender. This choice does not imply any loss of generality. Given the character of the Italian university system it can be assumed that all the variables note above are for the most part more than indirectly linked to the two chosen ones.

With regard to the group of personal variables, age is strongly correlated to academic role, given the system of career progression in Italian universities. At the same time, level of specialization of individual professionals is also implicitly taken into consideration, since the analysis is conducted precisely by scientific-disciplinary sector. Every academic scientist in Italy is classified in a specific sector of research, generally very clearly defined in terms of specialization. The Italian academic system is specifically subdivided into 14 macro disciplinary areas and 370 scientific sectors. The analysis conducted here is concentrated on the 8 areas of a technological-scientific character[2], which in turn include 183 sectors. The analysis initiates at the level of single sectors, implying that the individuals under observation are homogenous in their level of scientific specialization.

The inclusion of the structural-environmental variables listed above was also considered. An analysis of distribution of personnel among academic roles by geographic area showed substantial homogeneity, which induces the exclusion of location-related variables from the investigation. In addition, it seems little relevant to consider teaching load in the analysis, since the legal framework of the Italian academic system establishes this load *a priori*. The variable of relative prestige of department or institution also appears of little significance, given the publicly regulated nature of the Italian university system. The characteristics of university degrees are legislated and the recruitment of teaching personnel is based on rigidly regulated national competitions, inducing a situation of great homogeneity among universities. Only in recent years have universities attained a certain financial autonomy, still insufficient to facilitate differences in reputation.

---

[2] Mathematics and information sciences; physical sciences; chemical sciences; earth sciences; biological sciences; medical sciences; agriculture and veterinary sciences; industrial and information engineering. Civil engineering and architecture, a final technical-scientific area from the national university order, was discarded from consideration because the SCI™ would not be sufficiently exhaustive in representing the research output of this area.

## 3.1 Data set

The data utilized for the output component of the model were taken directly from the ORP (Italian Observatory of Public Research), developed in the authors' home institute. This observatory extracts data on scientific literature from the SCI™ and applies procedures for disambiguation and identification of the exact origin of the publications[3]. It lists all scientific articles authored by Italian university personnel[4] holding a position as assistant, associate or full professor during the three years under consideration (2001 to 2003), in the technological-scientific university disciplinary areas (UDAs). The research personnel were identified by extraction from a database at the Ministry of Universities and Research[5] and number approximately 33,000 scientists from the sectors indicated by Table 1. The data show several interesting points. Firstly, of the total of scientific staff, assistant professors number more than for any other role: approximately 38%, compared to 33.4% for associate professors and 28.8 for full professors. This division by role is quite different when men and women are observed separately. The latter, which represent slightly more than one quarter of the population, are much more concentrated in lesser roles. More than 55% of women fall within the role of assistant professor, compared to only 32% of men. The inverse situation occurs for senior roles – for each female full professor there are more than 8 males. This discrepancy, substantially consistent among all disciplinary areas, has historic-social causes and reflects the state of progress of feminine emancipation in education and employment.

[Table 1]

To permit the analysis of scientific production from single individuals over the period of observation, scientists who did not hold a staff role throughout the entire triennium were cut from the initial data set, eliminating all those who were assumed after December 31, 2000 or exited prior to January 1, 2003. Scientists that changed scientific disciplinary sector (SDS) for whatever reason were also cut from the data[6]. Those who changed their role within an SDS during the triennium due to career advancement were attributed with their role in the final year of observation (2003). The final data set used in the analysis is presented in Table 2. Comparing to Table 1, one sees a distortion in favour of senior roles (associate professors and especially full professors), caused by the procedures noted.

[Table 2]

## 3.2 Performance indicators

Individual performance was evaluated on the basis of a number of indicators:

---

[3] For an exhaustive description of the development and function of the observatory and the listings of scientific production by name for Italian university researchers see Abramo et al. (2007).

[4] Each indexed artiche is assigned to all the co-authors, regardless of their position in the listing.

[5] http://cercauniversita.cineca.it/php5/docenti/cerca.php. It was impossible to obtain direct information on age, gender, or marriage of individuals due to privacy regulations. Identification of gender was obtained by analysis of first names.

[6] Problems with homonymy in names would have made exact identification of individuals difficult as they moved from sector to sector, contributing potential errors in attribution, listing and count of publications.

- Output (O): the sum of publications realized by the scientist in the triennial under consideration.
- Fractional Output (FO): the sum of the scientist's contributions to the publications realized, the contribution for each publication being considered as the inverse of the number of co-authors.
- Contribution Intensity (CI): the ratio of FO to output. A value close to 1 indicates that the scientist generally excludes collaboration, publishing articles alone; the inverse, a value close to 0, indicates that the scientist tends to publish in co-authorship with many other colleagues.
- Scientific Strength (SS): equals the weighted sum of publications realized by the scientist. The weight is the normalized impact factor[7] of the publishing journal.
- Fractional Scientific Strength (FSS): analogous to FO but based on the scientific strength.
- Quality Index (QI): the ratio of scientific strength to output, indicating the average quality of the publications authored by the scientist.

The authors are aware of limitations arising from some of the methodological assumptions used. In particular, in this analysis only scientific journal publications are taken into consideration as research output, which excludes other codified forms of outputs such as proceedings, monographs, patents or prototypes. However in the scientific sectors taken into consideration, journal publications are actually highly representative of real output from research activity. It should be noted that when Italian universities submitted their products for consideration in the first national research evaluation (VTR-CIVR[8], 2006), journal articles were a minimum of 85% and a maximum of 99% of the total products selected by each university. In 7 of the 8 discipline areas under consideration, journal publications exceeded 90% of the total product submitted.

Overall, the most critical consideration is the correct quantification and classification of the articles for each university. In this regard, other than errors and limitations attributable to the data source[9], there may also be those arising from the identification of scientific production by author and institute name. But, as indicated by Abramo et al. (2007), the errors in the disambiguation process for author name do not induce substantial losses in significance for the analysis, due to the limited extent of such errors (2%) and to their uniform distribution among the data sets of the analytical model adopted.

A further critique could concern the use of the impact factor for the journal as a proxy for the publication quality, and therefore for the scientist's production. This assumption imposes a bias, but the bias diminishes at the moment that citations of single articles are considered (as amply described and analyzed in publications such as Weingart, 2004; Moed, 2002), and in the judgment of the authors the assumption does not significantly alter the study or the conclusions to which it gives rise.

---

[7] The distribution of the impact factors of journals is observed to differ substantially from sector to sector. The normalization of each journal's impact factor with respect to the sector average permits limiting the distortions embedded in comparing performances between different sectors.

[8] Triennial evaluation (2001-2003) of research activity in universities and major public research institutions; for details see http://vtr2006.cineca.it/

[9] The SCI™ lists approximately 4,800 international journals, which represent only a sample of the global scientific press. It also lacks uniform representation from the disciplines, for example being greatly weighted towards the life sciences.

With regards to assumptions concerning input, the major limitation lies in the impossibility to quantify the time dedicated to research activity by university professionals over the period under consideration. Further, there is no information on the frequency or duration of maternity time for women[10] or of sick leave in general. Although there is no reason to expect any gender diversity in distribution of illness, the negative impact of maternity on productivity by women could be notable, especially for assistant professors, where the average age is 43.

## 4. Results

In the triennial under observation, over 61.5% of Italian academic research personnel participated in at least one scientific publication listed in the SCI™ (Table 3). There is no significant difference in the data respecting men and women; 38.6% of the latter result as inactive, versus 38.5% among men. However, when the data are disaggregated by role and re-considered, this slight difference is overturned due to the fact that women are primarily concentrated in "lesser" or "less active" roles, as previously illustrated in Table 2. Among full professors, woman show 1.1% more activity than men. Among associate professors the difference in favour of women rises to 1.5% and among assistant professors as high as 5.2%.

When sorted by disciplinary areas the data do not show any strong lack of homogeneity between the sexes (Table 4). The maximum and minimum values refer respectively to male full professors in chemical sciences (where only 8.2% fail to produce any publications during the triennium) and to female associate professors in mathematics and information sciences (slightly less than two thirds fail to realize any publications in the triennial under observation). Independently of their role, women consistently result as more active than male colleagues in the areas of medical sciences, agriculture and veterinary sciences, and earth sciences. The opposite is true for the areas of industrial and information engineering, chemical sciences, physical sciences, and mathematics and information sciences. In addition, it is evident that the percentage of scientists that result as active in a specific area is correlated to the average intensity of publication in the area itself (Figure 1)[11].

[Table 3]
[Table 4]

Regarding only those scientists who publish, analyses at the aggregate level shows a noticeably skewed distribution in frequency of output (Figure 2). A 38% share of the active scientists produces less than one article per year over the arc of the triennium. The more productive scientists, however, contribute a notable portion of the scientific production: 20% of scientists realize over 53% of the scientific production in the whole of the areas under consideration for the national academic system. Significant and notable data also emerge concerning the difference between the sexes - the curve for distribution of publication frequency by women is more tapered than that for men, as can be seen from the inversion of the bars in the graph shown in Figure 2 (the skewness is 2.25 for women and 3.46 for men). The same occurs at single professional role level (Figure 3-5). Although with few differentiations, findings show that in general men are more concentrated than woman in the top-productivity ranks in each role.

---

[10] Under current law the normal duration is five months, but longer leaves are frequent and are permitted by the provisions. Leave is also permitted for men but they rarely take advantage of such provisions.

[11] Certain areas result as more productive than others either for internal reasons (time to complete projects and develop results is substantially less) or external reasons (the number of journals listed by the SCI™ in the area is larger than in others).

[Figure 1]
[Figure 2]
[Figure 3]
[Figure 4]
[Figure 5]

The analysis of the homogeneity of scientific production in different professional roles is illustrated in Table 5, which presents indexes of concentration for both genders in all three roles. The indexes used are the cumulative production of the first two deciles and the Gini coefficient. Considering each role, both the Gini coefficient and the cumulative production of the first and second deciles are consistently higher for men than for women. This phenomenon could be due, at least in part, to the presence of a relatively high number of "star" scientists among the male population.

[Table 5]

## 4.1 Differences in performance

The data presented in Table 6 now permit us to address the research questions posed at the origin of this study concerning the potential presence of significant differences in performance between men and women. For each indicator considered, the table reports the average general performance ($\bar{P}_{gk}$)[12], by gender and role. Table 6 also shows the average percentile rank (Rank %) for all the sectors included in the field of observation. These rank totals derive from the simple aggregation of the rank data for males and females in each sector. Basing analyses on the linearization of the data along an invariant scale from 0 to 100, the table permits evaluation of performance by individuals, independent of the sector in which they operate (i.e. independent of the number of scientists falling in the sector and of the sector's fertility in publications) In particular, Table 6 reports average data for men and women according to their role and the performance indicators considered. Higher performance for men can be observed along all dimensions of the evaluation. Notably, the overall average output per male scientist is 16.8% superior to that of female scientists. But the average quality of

---

[12] Average general performance ($\bar{P}_{gk}$) of scientists of gender "g" and role "k" is calculated as:

$$\bar{P}_{gk} = \frac{1}{Staff_{gk}} \sum_{j=1}^{n_{SDS}} \frac{\bar{Y}_{gjk}}{\bar{Y}_{jk}} \cdot Staff_{gjk}$$

$\bar{Y}_{gjk}$ = average performance of scientists of gender "g" and role "k", in sector "j"

$\bar{Y}_{jk}$ = average performance of scientists of role "k", in sector "j"

$Staff_{gjk}$ = number of scientists of gender "g" and role "k", in sector "j"

$Staff_{gk}$ = total of scientists of gender "g" and role "k"

$n_{SDS}$ = total of scientific sectors

production is only 4.5% superior (Quality Index, total in far right column of Table 6). Women also tend to collaborate more, seeing as their average contribution intensity (CI) is 6% inferior to that of men. The spread between the qualitative productivity indicators (SS and FSS) is obviously greater due to the combined effect of the differences found above.

[Table 6]

Analysis by role shows that, in terms of output, the average general production of men is greater than that of women in all of the three roles considered: +13.3% for full professors, +12.3% for associate professors and +17.5% for assistant professors. When considering the qualitative dimension of scientific production, the performance difference seems to increase – the general average scientific strength (SS) of men is 19.7% greater than that of women among full professors, 15.9% greater among associate professors and 20.2% greater among assistant professors.

Finally, the data also indicate a certain difference between the sexes relative to contribution, with a spread of once again in favour of men. In terms of contribution intensity (CI), the men's figures are greater than that of women by 2.9% among full professors, 5% among associate professors and 4.4% among assistant professors. The data for average percentile rank of the two sexes can be superimposed on the consideration of average performance. In particular, the percentile rank compresses the difference, reducing the weight of particularly anomalous data such as those that may be attributed to star scientists, but the overall higher performance of males still rests unchanged for all the roles and indicators considered.

It can also be noted that the difference in average percentile rank between men and women generally tends to decrease with increased stature of professional role. Looking at output (O), the difference in percentile rank between genders is 4.8% among assistant professors, 3% among associate professors, and only 1.5 per cent among full professors. The same trend can be seen with other indexes of observation. However, for the role of assistant professor, the difference in performance may have been amplified by not having taken into account the probable occurrence of maternity leaves.

As an alternative means of examination, the difference in performance between men and women was also calculated by applying the casual variables sequence criterion to the entire active population. Beginning with the performance ranking of each male scientist in his discipline sector, the distance between the ideal and effective cases was measured:

$$R^{diff}_{M-j} = R^{\max}_{M-j} - R^{eff}_{M-j} \qquad [1]$$

where:

$R^{\max}_{M-j}$ = sum of the ranks of males in sector j under the hypothesis of maximum differentiation*

$R^{eff}_{M-j}$ = sum of the ranks of males in sector j

*"maximum differentiation" is understood as the situation in which the highest performing woman is still ranked below the lowest performing male*

The value $R^{diff}_{M-j}$ therefore represents the "distance" for the ideal situation of maximum performance difference between genders in favour of males. The same calculation is completed for women, and through comparison between $R^{diff}_{M-j}$ and $R^{diff}_{F-j}$, it can be determined which of the two populations, male or female, obtains a higher overall ranking. The simple sum of the data by sector provides the overall view at the level of discipline area.

$$R^{diff}_{M-A} = \sum_{j=1}^{n_A}(R^{max}_{M-j} - R^{eff}_{M-j})  \quad [2]$$

where:
$R^{diff}_{M-A}$ = distance from the situation of maximum differentiation for area A
$n_A$ = number of sectors included in area A

This analysis once again gives a comparison between $R^{diff}_{M-A}$ and $R^{diff}_{F-A}$, indicating which of the two populations, male or female, obtains a higher average overall ranking at the level of discipline area (Tables 7, 8, 9). Although men do have an overall higher performance than that of women (see the last row of each table), the contrary occurs in some specific areas: for full and associate professors in the industrial and information engineering area and for full professors and assistant professors in agriculture and veterinary sciences. Female full professors result as having higher output than men in the physical and chemical sciences areas, but for scientific strength in the latter area it is males that obtain a higher performance. For the physical sciences, a similar inversion in favour of males occurs when examining the fractional dimension of performance.

The case of agriculture and veterinary sciences is singular – women shine in the roles of full professor and assistant professor, but they are unseated by men in the intermediate role of associate professor. In earth sciences, women assistant professors seem stronger than their male colleagues. With the advancement of professional role the situation changes – among associate professors men achieve the higher performance in 5 indicators out of 6; among full professors men score better in all 6 indexes.

[Table 7]
[Table 8]
[Table 9]

**4.2 Analysis at the level of single sectors**

Moving down to the level of sectors, there are further interesting points for consideration. Tables 10, 11 and 12 indicate, for each area, the number of sectors in which the average percentile rank of women is not inferior to that of men. The tables also indicate, in parentheses, the weight of each sector in terms of the total number of professionals in the area. It should be noted that the comparison is only possible in the sectors where both males and females hold professional roles. Since the representation of women varies among professional roles there is also variation in the total number of sectors per role in which a comparison between is possible: 110 sectors for full professors, 146 for associate professors and 147 for assistant professors.

Considering the population of full professors first (Table 10), the average percentile rank for output by women is not less than that of men in 43 sectors out of 110 (39.1%) and 28.6% of the professors are employed in those sectors. Very similar indications are obtained for other performance indexes, while important differentiations emerge at the level of disciplinary area. For agriculture and veterinary sciences, still referring to the role of full professor, women demonstrate performance not less than men in 14 sectors out of 20 in terms of output, 11 out of 20 if considering scientific strength (SS) and 13 out of 20 considering fractional output (FO). The case of physical sciences also presents an interesting situation. Women register productivity data that are not inferior to those of men in four sectors out of five, in terms of output and scientific strength. However, when the fractional dimension is considered, the difference between men and women is inverted in the sector PHY/01 (experimental physics), which is the most populous sector (38% of all scientists in the area fall within this sector). Still with regards to full professors, it can be noted that industrial and information engineering, an area where the representation of women is truly marginal (see Table 1: one female professor for every 25 males), women register a performance which is not inferior to that of men in six sectors out of 14 for output, and five out of 14 for scientific strength. Finally, in medical sciences (an area which represents 31% of the entire field of investigation, in terms of number of professionals), comparison indicates a performance by women which is not inferior to that of men in 10 sectors of 28 for scientific strength and 11 of 28 for output. In the mathematical sciences area, the average percentile rank for males is not less than that of women in 6 out 7 cases, for output and scientific impact. However, in terms of contribution intensity, it can be seen that performance by women is not lesser in 4 sectors out of 7.

[Table 10]
[Table 11]
[Table 12]

Finally, an examination conducted on the data within the sectors reveals the absence of correlation between the numbers of women and their general ranking with respect to male colleagues in each sector.

Comparative analysis for the data in Tables 10, 11 and 12 offers interesting highlights concerning variability of performance differentials with respect to professional role. The realities that emerge are not completely consistent from one indicator to the next. For example, a review of output would seem to suggest that the gap between men and women tends to lessen with increased professional status. For assistant professors, the performance of women is not inferior to that of men in 49 sectors out of 147 (33.7 of cases), compared to 55 out of 146 (37.7%) for associate professors, and as previously indicated, 43 out of 110 (39.1%) for full professors. However, when the qualitative dimension is considered, there is no evidence that the gap varies with role. For example, in scientific strength, the performance of women remains quite uniform – it is not inferior to that of men in 58 sectors out of 147 for assistant professors (39.5% of cases), in 55 sectors out of 146 for associate professors (37.7%) and in 42 out of 100 sectors for full professors (38.2%). Contribution intensity also seems to flatten the variability by role in performance gap between men and women. In general, it is in the population of associate professors that the higher performance of males is challenged in the greatest number of sectors. With career progress, the sectorial gaps between the sexes vary greatly from discipline to discipline. For example, in earth sciences, the number of sectors in which performance by males registers not less than that of women actually seems to increase with career progression.

This occurs in almost every sector for full professors (seven out of eight for output, and six out of eight for scientific strength), but only in 6 out of 12 sectors for associate professors and 6 to 7 out of 11 for assistant professors.

## 5. Final considerations

The analysis of productivity differences between men and women employed in research has always attracted interest among science sociologists, whose studies agree in acknowledging a higher performance among men than women. The study reported here, analyzing the technological-scientific disciplines of the entire Italian academic system, confirms the existing literature but also brings to light significant differences in the distribution of performance between the sexes. Males do demonstrate a higher average productivity with respect to that of females for all the performance indicators considered. However one of the new and interesting facts is that it is above all in the quantitative dimension of output where the major gap is recorded. In terms of quality index and contribution intensity, the gap between the sexes, though still present, seems less pronounced. The performance gap also seems to reduce with career advancement. This could in part be due to the effect of maternity, it being reasonable that the experience of motherhood would be more frequent, for age reasons, among the lesser university career roles. This result seems to coincide with the conclusions by Stack (2004), whose work suggests that women with preschool aged children publish less than others. In effect, the average age of female research professionals in the Italian academic system for the period under observation is 43 years, falling within the final family life phase for the presence of very young children.

Although this study does not permit proceeding to inter-temporal comparisons, the average gap revealed, while still significant, is notably reduced compared to results reported by other authors. This lends value to the increasingly common thesis of a progressive reduction over time for the performance gap between the sexes, as proposed by Cole and Zuckerman (1984), Xie and Shauman (1998) and Leahey (2006).

Adding to current literature, the study reported also highlights that there are important sectorial specificities in the differences between the sexes, but these generally do not raise any challenge to the higher performance of men in all dimensions of the evaluation of scientific performance. But if it is true that the average performance of men is higher than that of women, this is not the case in all sectors of research professionals. In terms of output by full professors, for 43 sectors out of 100 women do not demonstrate any lesser performance than that of men. For associate professors this occurs in 55 sectors out of 146 and for assistant professors in 49 out of 147. Further, certain areas result as particularly interesting for interpretation of the gender gap. In industrial and information engineering for example, the feminine presence is truly marginal, representing 10% of scientists, and among full professors only 4% of the total. Yet women in this area demonstrate performance not less than their male colleagues in just less than half of the sectors. This could suggest further questions revolving around the hypothesis of discrimination between sexes in this area in particular and in the entire academic system in general. In partial contrast to tendencies in the literature, certain results emerge concerning non-productive professionals. For each role, the percentage of non-productive males is higher than that for females, while the reverse is true in the overall population, though with a minimal difference. However women show a higher concentration than men in the lowest levels of productivity. The contrary situation registers for the highest levels of performance. It is therefore possible that males are characterized by a higher concentration of "star scientists", and this, with all probability, would play a significant role in the generally higher performance of men than women. The authors will attempt to respond to this suggestion in future work.

|  |  | Industrial and information engineering | Agriculture and veterinary sciences | Biological sciences | Chemical sciences | Earth sciences | Physics | Mathematics and information sciences | Medical sciences | Total |
|---|---|---|---|---|---|---|---|---|---|---|
| Full Professors | M | 1,488 (96%) | 865 (91%) | 1,099 (77%) | 832 (88%) | 350 (91%) | 748 (94%) | 793 (85%) | 2,260 (92%) | 8,435 (89%) |
|  | F | 60 (4%) | 90 (9%) | 331 (23%) | 109 (12%) | 35 (9%) | 47 (6%) | 144 (15%) | 188 (8%) | 1,004 (11%) |
| Associate Professors | M | 1,299 (90%) | 660 (74%) | 859 (54%) | 823 (70%) | 371 (77%) | 788 (84%) | 678 (62%) | 2,684 (81%) | 8,162 (75%) |
|  | F | 148 (10%) | 231 (26%) | 729 (46%) | 360 (30%) | 113 (23%) | 146 (16%) | 414 (38%) | 643 (19%) | 2.784 (25%) |
| Assistant Professors | M | 1,115 (82%) | 646 (58%) | 747 (40%) | 481 (47%) | 274 (65%) | 572 (73%) | 572 (53%) | 3,249 (68%) | 7,656 (62%) |
|  | F | 241 (18%) | 470 (42%) | 1,102 (60%) | 544 (53%) | 148 (35%) | 215 (27%) | 507 (47%) | 1,548 (32%) | 4,775 (38%) |
| Total | M | 3,901 (90%) | 2,171 (73%) | 2,705 (56%) | 2,136 (68%) | 995 (77%) | 2,108 (84%) | 2,043 (66%) | 8,194 (77%) | 24,253 (74%) |
|  | F | 449 (10%) | 791 (27%) | 2,162 (44%) | 1,013 (32%) | 296 (23%) | 408 (16%) | 1,065 (34%) | 2,379 (23%) | 8,563 (26%) |
| Total, both sexes |  | 4,350 | 2,962 | 4,867 | 3,149 | 1,291 | 2,516 | 3,108 | 10,573 | 32,816 |

*Table 1: Distribution of Italian university staff by gender, role and disciplinary area: average of data for years 2001 to 2003*

|  | Full Professors | Associate Professors | Assistant Professors | Total |
|---|---|---|---|---|
| Male | 8,686 (88.7%) | 7,596 (73.4%) | 5,401 (60.7%) | 21,683 (74.7%) |
| Female | 1,102 (11.3%) | 2,758 (26.6%) | 3,493 (39.3%) | 7,353 (25.3%) |
| Total | 9,788 (33.7%) | 10,354 (35.76%) | 8,894 (30.6%) | 29,036 |

*Table 2: Distribution of Italian university research staff by gender and role; data set 2001 to 2003*

|  | Full Professors | Associate Professors | Assistant Professors | Total |
|---|---|---|---|---|
| Male | 5,895 (67.9%) | 4,526 (59.6%) | 2,921 (54.1%) | 13,342 (61.5%) |
| Female | 760 (69.0%) | 1,684 (61.1%) | 2,071 (59.3%) | 4,515 (61.4%) |
| Total | 6.655 (68.0%) | 6,210 (60.0%) | 4,992 (56.1%) | 17,857 (61.5%) |

*Table 3: Distribution of scientists who publish, by gender and role*

|  |  | Industrial and information engineering | Agriculture and veterinary sciences | Biological sciences | Chemical sciences | Earth sciences | Physical sciences | Mathematics and information sciences | Medical sciences | Total |
|---|---|---|---|---|---|---|---|---|---|---|
| Full Professors | M | 52.7% | 50.6% | 80.9% | 91.8% | 56.0% | 72.9% | 56.5% | 73.8% | 67.9% |
|  | F | 50.7% | 61.0% | 72.3% | 87.6% | 62.9% | 66.0% | 52.5% | 76.6% | 69.0% |
| Associate Professors | M | 52.3% | 45.9% | 69.4% | 83.9% | 48.8% | 62.3% | 44.4% | 60.7% | 59.6% |
|  | F | 56.0% | 50.0% | 71.2% | 81.3% | 52.1% | 56.9% | 37.1% | 61.3% | 61.1% |
| Assistant Professors | M | 51.4% | 43.8% | 69.5% | 86.4% | 42.6% | 72.9% | 46.5% | 47.8% | 54.1% |
|  | F | 48.7% | 52.0% | 68.4% | 81.6% | 50.0% | 66.1% | 40.7% | 55.3% | 59.3% |
| Total | M | 52.3% | 47.5% | 74.7% | 87.8% | 50.3% | 68.9% | 50.2% | 60.6% | 61.5% |
|  | F | 52.0% | 52.7% | 70.2% | 82.3% | 52.7% | 62.5% | 41.1% | 59.6% | 61.4% |
|  | Total | 52.2% | 48.9% | 72.8% | 86.1% | 50.8% | 67.9% | 47.1% | 60.4% | 61.5% |

*Table 4: Percentage of scientists who publish, by gender, role and discipline area*

|  | Full Professor | | Assistant Professor | | Assistant Professors | |
|---|---|---|---|---|---|---|
|  | M | F | M | F | M | F |
| Gini coefficient | 0.509 | 0.448 | 0.488 | 0.449 | 0.472 | 0.437 |
| Cumulative production 1st decile | 47.5% | 34.0% | 30.4% | 19.9% | 22.4% | 13.1% |
| Cumulative production 2nd decile | 65.7% | 56.9% | 48.4% | 40.1% | 39.8% | 29.0% |

*Table 5: Indexes of concentration of scientific production by gender and role; data from 2001 to 2003 for scientists who publish*

| Index | Gender | Full Professors | | Assistant Professors | | Assistant Professors | | Total |
|---|---|---|---|---|---|---|---|---|
|  |  | $\overline{P}_{gk}$ | Rank % | $\overline{P}_{gk}$ | Rank % | $\overline{P}_{gk}$ | Rank % | $\overline{P}_{gk}$ |
| O | M | 1.252 (+13.3%) | 63.0 | 0.94 (+12.3%) | 55.4 | 0.848 (+17.5%) | 53.4 | 1.032 (+16.8%) |
|  | F | 1.105 | 61.5 | 0.837 | 52.4 | 0.722 | 48.6 | 0.884 |
| SS | M | 1.286 (+19.7%) | 56.8 | 0.939 (+15.9%) | 48.2 | 0.828 (+20.2%) | 46.6 | 1.038 (+21.8%) |
|  | F | 1.074 | 55.3 | 0.81 | 45.9 | 0.689 | 42.4 | 0.853 |
| FO | M | 1.238 (+15.7%) | 56.9 | 0.965 (+16.5%) | 49.9 | 0.871 (+25.0%) | 47.3 | 1.042 (+21.3%) |
|  | F | 1.07 | 55.0 | 0.828 | 46.3 | 0.697 | 41.9 | 0.860 |
| FSS | M | 1.287 (+22.6%) | 56.4 | 0.956 (+23.0%) | 48.9 | 0.846 (+27.6%) | 46.8 | 1.049 (+27.5%) |
|  | F | 1.05 | 54.4 | 0.777 | 45.5 | 0.663 | 42.0 | 0.823 |
| QI | M | 1.03 (+2.7%) | 53.0 | 0.986 (+1.4%) | 48.6 | 0.994 (+1.7%) | 49.5 | 1.005 (+4.5%) |
|  | F | 1.003 | 51.2 | 0.972 | 48.5 | 0.977 | 48.3 | 0.961 |
| CI | M | 0.983 (+2.9%) | 50.6 | 1.032 (+5.0%) | 52.4 | 1.021 (+4.4%) | 51.5 | 1.010 (+6.0%) |
|  | F | 0.955 | 49.0 | 0.983 | 50.5 | 0.978 | 48.8 | 0.953 |

*Table 6: Average general performance ($\overline{P}_{gk}$) and average percentile rank (Rank %) of men and women by role (percentage differences indicated in brackets)*

| Area | O | SS | FO | FSS | QI | CI |
|---|---|---|---|---|---|---|
| Industrial and information engineering | F | F | F | F | F | F |
| Agriculture and veterinary sciences | F | F | F | F | M | F |
| Biological sciences | M | M | M | M | M | M |
| Chemical sciences | F | M | F | M | M | M |
| Earth sciences | M | M | M | M | M | M |
| Physical sciences | F | F | M | M | M | M |
| Mathematics and information science | M | M | M | M | M | F |
| Medical sciences | M | M | M | M | M | M |
| Total | M | M | M | M | M | M |

*Table 7: Analysis of the average position of male and female full professors using the casual variables sequence criterion*

| Area | O | SS | FO | FSS | QI | CI |
|---|---|---|---|---|---|---|
| Industrial and information engineering | M | F | F | F | F | F |
| Agriculture and veterinary sciences | M | M | M | M | M | F |
| Biological sciences | M | M | M | M | F | M |
| Chemical sciences | F | F | M | F | F | M |
| Earth sciences | M | M | M | M | F | M |
| Physical sciences | F | F | F | F | M | F |
| Mathematics and information sciences | M | M | M | M | M | M |
| Medical sciences | M | M | M | M | F | M |
| Total | M | M | M | M | F | M |

*Table 8: Analysis of the average position of male and female associate professors using the casual variables sequence criterion*

| Area | O | SS | FO | FSS | QI | CI |
|---|---|---|---|---|---|---|
| Industrial and information engineering | M | M | M | M | M | M |
| Agriculture and veterinary sciences | M | F | F | F | F | F |
| Biological sciences | M | M | M | M | M | M |
| Chemical sciences | M | M | M | M | M | M |
| Earth sciences | F | F | F | F | F | M |
| Physical sciences | M | M | M | M | F | F |
| Mathematics and information science | M | M | M | M | M | M |
| Medical sciences | M | M | M | M | F | M |
| Total | M | M | M | M | M | M |

*Table 9: Analysis of the average position of male and female assistant professors using the casual variables sequence criterion*

| Area | # SDS | O | SS | FO | FSS | QI | CI |
|---|---|---|---|---|---|---|---|
| Industrial and information engineering | 14 | 6 (33.8%) | 5 (31.7%) | 7 (36.9%) | 5 (24.8%) | 6 (24.7%) | 7 (41.2%) |
| Agriculture and veterinary sciences | 20 | 14 (59.5%) | 11 (48.6%) | 13 (56.6%) | 11 (50.8%) | 8 (32.2%) | 10 (43.4%) |
| Biological sciences | 18 | 4 (10.4%) | 6 (26.2%) | 4 (12.5%) | 5 (26.0%) | 9 (57.0%) | 4 (27.1%) |
| Chemical sciences | 10 | 2 (23.1%) | 3 (39.5%) | 3 (45.0%) | 2 (23.1%) | 4 (42.3%) | 5 (54.7%) |
| Earth sciences | 8 | 1 (14.0%) | 2 (22.3%) | 1 (14.0%) | 1 (14.0%) | 3 (23.9%) | 2 (25.0%) |
| Physical sciences | 5 | 4 (70.9%) | 4 (70.9%) | 3 (32.9%) | 3 (32.9%) | 2 (22.0%) | 1 (10.9%) |
| Mathematics and information sciences | 7 | 1 (4.0%) | 1 (4.0%) | 1 (4.0%) | 1 (4.0%) | 2 (20.9%) | 4 (62.8%) |
| Medical sciences | 28 | 11 (27.8%) | 10 (22.2%) | 10 (24.6%) | 10 (22.2%) | 10 (27.6%) | 12 (24.0%) |
| Total | 110 | 43 (28.6%) | 42 (31.1%) | 42 (27.9%) | 38 (24.7%) | 44 (33.7%) | 45 (34.0%) |

*Table 10: Number of technical-scientific sectors in which the average percentile rank of female full professors is not inferior to that of males (figures in brackets indicate weight of sectors in terms of percentage of total professionals in the area)*

| Area | # SDS | O | SS | FO | FSS | QI | CI |
|---|---|---|---|---|---|---|---|
| Industrial and information engineering | 24 | 13 (41.6%) | 14 (44.6%) | 14 (56.2%) | 14 (54.1%) | 14 (48.9%) | 12 (52.7%) |
| Agriculture and veterinary sciences | 26 | 7 (16.6%) | 7 (17.6%) | 10 (27.0%) | 8 (25.0%) | 12 (39.6%) | 15 (59.7%) |
| Biological sciences | 19 | 5 (28.2%) | 6 (30.4%) | 4 (27.0%) | 6 (30.4%) | 9 (43.8%) | 7 (49.4%) |
| Chemical sciences | 11 | 4 (31.6%) | 4 (31.6%) | 4 (31.6%) | 4 (31.6%) | 4 (50.7%) | 6 (39.9%) |
| Earth sciences | 12 | 6 (41.0%) | 6 (52.4%) | 7 (46.6%) | 5 (50.1%) | 5 (45.9%) | 6 (47.3%) |
| Physical sciences | 8 | 4 (63.7%) | 3 (52.8%) | 4 (63.7%) | 3 (52.8%) | 2 (45.0%) | 6 (74.9%) |
| Mathematics and information sciences | 9 | 2 (17.8%) | 1 (13.8%) | 1 (4.0%) | 0 (-) | 4 (36.3%) | 2 (8.2%) |
| Medical sciences | 37 | 14 (32.7%) | 18 (46.9%) | 10 (20.2%) | 16 (43.9%) | 24 (68.4%) | 17 (42.9%) |
| Total | 146 | 55 (33.4%) | 59 (37.7%) | 54 (30.9%) | 56 (37.3%) | 74 (52.2%) | 71 (46.3%) |

*Table 11: Number of technical-scientific sectors in which the average percentile rank of female associate professors is not inferior to that of males; figures in brackets indicate weight of sectors in terms of percentage of total professionals in the area*

| Area | # SDS | O | SS | FO | FSS | QI | CI |
|---|---|---|---|---|---|---|---|
| Industrial and information engineering | 23 | 9 (22.4%) | 10 (30.3%) | 12 (29.0%) | 11 (26.4%) | 12 (35.5%) | 12 (43.0%) |
| Agriculture and veterinary sciences | 27 | 12 (44.5%) | 13 (49.5%) | 15 (55.4%) | 13 (44.7%) | 16 (59.2%) | 16 (52.9%) |
| Biological sciences | 19 | 6 (14.8%) | 7 (19.6%) | 6 (28.2%) | 5 (25.7%) | 5 (15.5%) | 6 (37.8%) |
| Chemical sciences | 11 | 3 (15.3%) | 4 (16.1%) | 4 (20.4%) | 4 (20.4%) | 7 (42.1%) | 3 (26.4%) |
| Earth sciences | 11 | 4 (38.6%) | 5 (50.4%) | 4 (35.1%) | 4 (41.4%) | 6 (64.2%) | 4 (35.1%) |
| Physical sciences | 7 | 3 (19.9%) | 3 (19.9%) | 3 (19.9%) | 3 (19.9%) | 2 (45.0%) | 4 (69.3%) |
| Mathematics and information sciences | 8 | 1 (16.9%) | 2 (30.7%) | 1 (16.9%) | 1 (16.9%) | 2 (30.7%) | 3 (53.0%) |
| Medical sciences | 41 | 11 (16.2%) | 14 (35.9%) | 9 (14.1%) | 11 (17.1%) | 23 (55.6%) | 14 (30.9%) |
| Total | 147 | 49 (19.7%) | 58 (29.4%) | 54 (23.5%) | 52 (23.1%) | 73 (42.2%) | 62 (39.6%) |

*Table 12: Number of technical-scientific sectors in which the average percentile rank of female assistant professors is not inferior to that of males; figures in brackets indicate weight of sectors in terms of percentage of total professionals in the area*

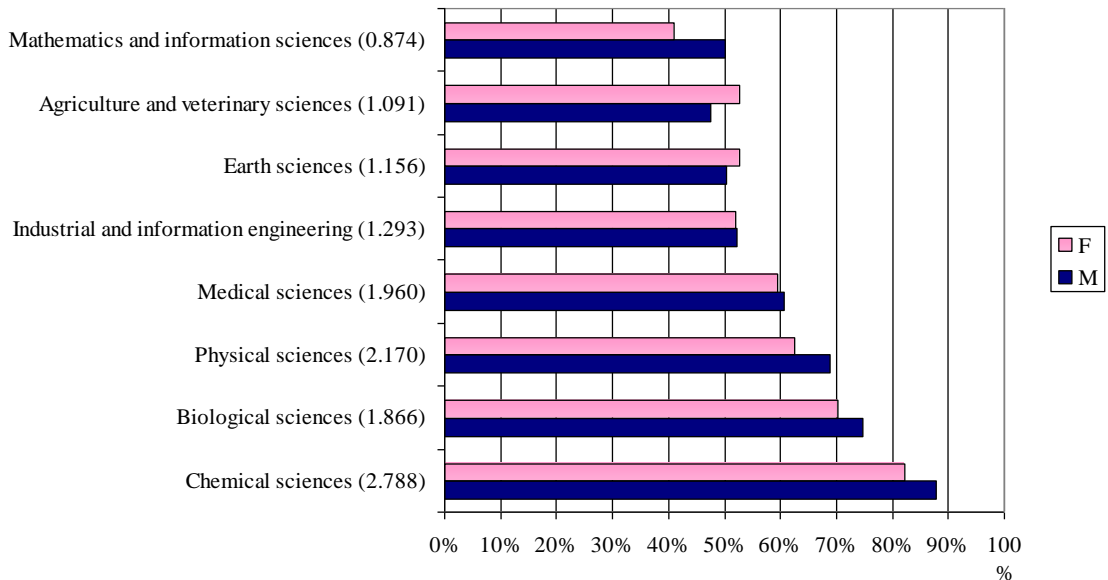

*Figure 1: Percentage of scientists who publish, by gender and discipline area. The intensity of publication is indicated in parentheses (average annual publications per active professional).*

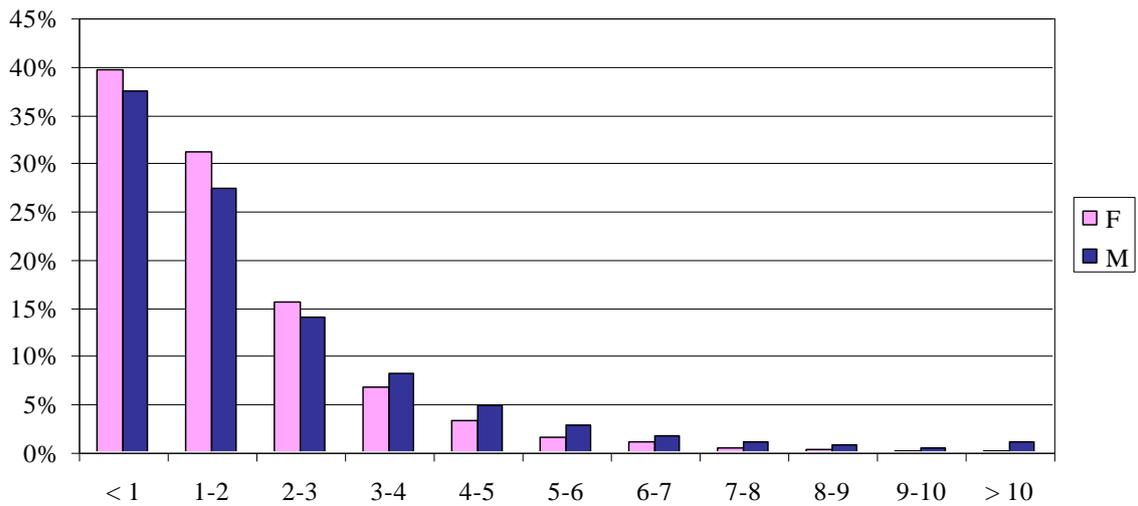

*Figure 2: Average frequency of scientific production by Italian academic research personnel; data from 2001 to 2003 for scientists who publish.*

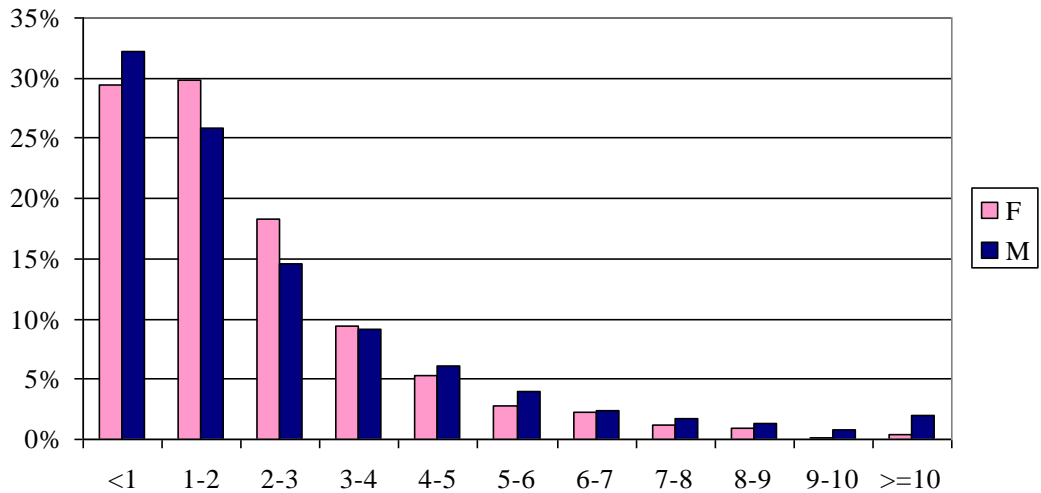

*Figure 3: Average frequency of scientific production by Italian academic research personnel; data from 2001 to 2003 for full professors who publish*

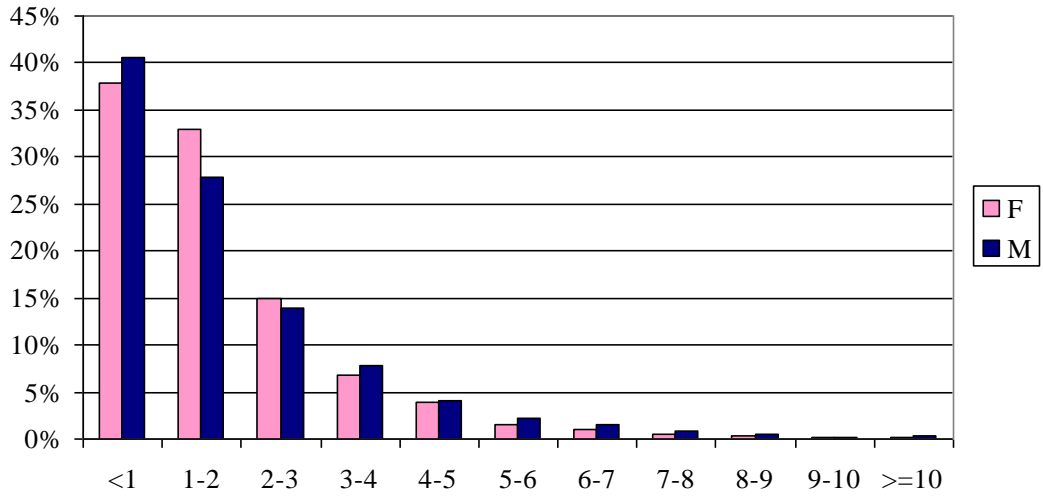

*Figure 4: Average frequency of scientific production by Italian academic research personnel; data from 2001 to 2003 for associate professors who publish*

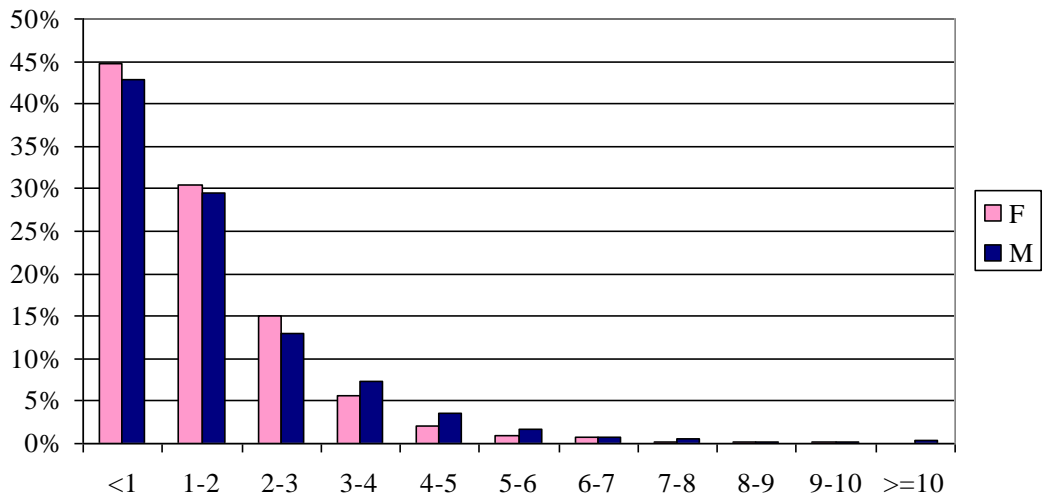

*Figure 5: Average frequency of scientific production by Italian research academic personnel; data from 2001 to 2003 for assistant professors who publish*